\documentclass[useAMS,usenatbib]{mn2e}

\usepackage{graphics}

\title[Angular momentum and mass evolution]
{Angular momentum and mass evolution of contact binaries}

\author[K. Gazeas and K. St\c{e}pie\'n]
{K. Gazeas$^{1}$\thanks{e-mail: kgaze@physics.auth.gr, kgazeas@cfa.harvard.edu} and K. St\c{e}pie\'n$^{2}$\thanks{e-mail: kst@astrouw.edu.pl} \\
$^{1}$Harvard-Smithsonian Center for Astrophysics, 60 Garden Street, Cambridge, MA 02138, USA\\
$^{2}$Warsaw University Observatory, Al. Ujazdowskie 4, 00-478 Warszawa, Poland}

\date{Accepted --.
         Received -- ;
         in original form --}

\pubyear{2007}

\begin{document}
\maketitle
\label{firstpage}

\begin{abstract}

Various scenarios of contact binary evolution have been proposed in the
past, giving hints of (sometimes contradictory) evolutionary sequence
connecting A-type and W-type systems. As the components of close detached
binaries approach each other and contact binaries are formed, following
evolutionary paths transform them into systems of two categories: A-type
and W-type.  The systems evolve in a similar way but under slightly
different circumstances. The mass/energy transfer rate is different,
leading to quite different evolutionary results. An alternative scenario of
evolution in contact is presented and discussed, based on the observational
data of over a hundred low-temperature contact binaries. It results from
the observed correlations among contact binary physical and orbital
parameters.  Theoretical tracks are computed assuming angular momentum loss
from a system via stellar wind, accompanied by mass transfer from an
advanced evolutionary secondary to the main sequence primary. Good
agreement is seen between the tracks and the observed graphs. Independently
of details of the evolution in contact and a relation between A-type and
W-type systems, the ultimate fate of contact binaries involves the
coalescence of both components into a single fast rotating star.

\end{abstract}

\begin{keywords}
stars: contact -- stars: eclipsing -- stars: binary -- stars: evolution
\end{keywords}

\section{Introduction}  
\label{sect:intro}

Based on mass-radius and color-luminosity diagrams \citet{Hilditch1988},
\citet{Hilditch1989}, as well as other investigators, suggest that W
UMa-type stars of A-type are more evolved than W-type. They developed an
original idea of \citet{Lucy1976}. \citet{Maceroni1996} and \citet{YE2005}
noted that this might be wrong, by studying the mass-luminosity and
period-angular momentum (AM) diagrams. Later on, \citet{GN2006} showed that
A-type systems cannot be more evolved than W-type, since their total mass
and total angular momentum are larger. It seems that A-type and W-type
systems belong to the same family of cool contact binaries, but the
evolutionary relations between both types may be more complicated than
hitherto thought.

Recently, \citet{Stepien2004} developed a model of a W~UMa-type binary with
the currently less massive component being the more evolved one. Such a
model is conceptually very close to that used to explain the semi-detached
Algols. In his model, the current secondary (less massive) components must
have a very low mass in some cases but possess small helium cores to
explain systems like AW~UMa or SX~Crv \citep{Paczynski2007}. In a new
model, the problem of thermal equilibrium of both components is solved by
assuming that contact binaries are past mass exchange with a mass ratio
reversal \citep{Stepien2006a}.

In this paper we discuss the physical and geometrical parameters of the
components of more than a hundred cool contact binaries. Our sample is
based mainly on the list of contact binaries given by
\citet{Kreiner2003}. Half of the systems from the list has solutions
published in the frame of the W~UMa project (papers I-VI)
\citep{Kreiner2003, Baran2004, Zola2004, Gazeas2005, Zola2005,
Gazeas2006}. Data for the remaining systems were collected from literature.
Only binaries with accurate solutions based on high quality photometric
light curves and good radial velocity curves for both components were
included. We show that several relations and correlations exist among the
discussed parameters. Some of them may be useful in the future for
approximate estimates of masses and radii of contact binaries for which
only orbital periods are known, which is a typical situation for binaries
detected in massive photometric surveys. Later we discuss the evolutionary
scenario for different types of binaries based on models suggested by
\citet{Stepien2004, Stepien2006a,Stepien2006b}.

In our study we consider component "1" as the presently more massive one. Our
assumption is based on the double-lined spectroscopic observations, where
the mass ratio is taken as $q=M_{\rmn{2}}/M_{\rmn{1}} \le 1$.

\section{Observed Properties of W~UMa-type stars}
\label{sect:main}

Cool contact binaries are divided into two categories according to the
relative minima depth \citep{Binnendijk1970}. Those with a primary minimum
being a transit are called A-type, while when it is an occultation
the binary is of W-type.  This can be translated into a temperature
difference: primary components are hotter than secondaries in A-type
binaries whereas the opposite occurs in W-type stars. A typical temperature
difference is of the order of a few hundred kelvin but there are many
binaries with the difference very close to zero and even some alternating
between A- and W-type. Existing models of energy transport between the
components always predict hotter primaries \citep{Lucy1968, Kahler2002,
Kahler2004}. So far, the only acceptable explanation for W-type effect
assumes the existence of cool, dark spots on primaries which results in a
drop of the surface averaged temperature \citep{Binnendijk1970, Eaton1980,
Stepien1980}. Equivalently, hot spots on secondaries can also do. A-type
binaries have, on average, longer orbital periods, earlier spectral types
(i. e. more massive primaries) and lower mass ratios compared to W-type but
a significant overlap is present.  It is not clear whether the division
into A-type and W-type binaries is superficial or the differences between
both types are deeper and more fundamental.

\begin{table*}
\caption{Results derived from the light curve modeling for 52 W-type contact binaries}
\begin{flushleft}
\begin{small}   
\begin{tabular}{lccccccccccc}
\hline
Name			&$P_{\rm{rot}}$	&logP		&$M_{1}$		&$M_{2}$	&q		&$R_{1}$	&$R_{2}$	&a		&$H_{\rm{orb}}$		&Type	&Ref	\\	
			&(days)		&		&$(M_{\odot})$		&$(M_{\odot})$&		&$(R_{\odot})$	&$(R_{\odot})$	&$(R_{\odot})$	&($10^{-51}$ gcm$^{2}$s$^{-1}$)   &	&	\\	
\hline	
	CC~Com		&	0.2211	&	-0.6554	&	0.690	&	0.360	&	0.522	&	0.682	&	0.507	&	1.563	&	1.832	&	W	&	7	\\	
	V523~Cas		&	0.2337	&	-0.6313	&	0.637	&	0.319	&	0.501	&	0.692	&	0.505	&	1.572	&	1.575	&	W	&	8	\\	
	RW~Com		&	0.2373	&	-0.6247	&	0.920	&	0.310	&	0.337	&	0.821	&	0.501	&	1.728	&	2.043	&	W	&	9	\\	
	44~Boo			&	0.2678	&	-0.5722	&	0.900	&	0.439	&	0.488	&	0.852	&	0.614	&	1.926	&	2.865	&	W	&	10	\\	
	VW~Cep		&	0.2783	&	-0.5555	&	0.897	&	0.247	&	0.275	&	0.924	&	0.515	&	1.875	&	1.715	&	W	&	7,11	\\	
	BX~Peg		&	0.2804	&	-0.5522	&	1.020	&	0.380	&	0.373	&	0.940	&	0.600	&	2.016	&	2.812	&	W	&	12	\\	
	XY~Leo		&	0.2841	&	-0.5465	&	0.870	&	0.435	&	0.500	&	0.874	&	0.637	&	1.987	&	2.823	&	W	&	13	\\	
	RW~Dor		&	0.2854	&	-0.5445	&	0.640	&	0.430	&	0.672	&	0.772	&	0.644	&	1.865	&	2.196	&	W	&	14	\\	
	BW~Dra		&	0.2923	&	-0.5342	&	0.891	&	0.250	&	0.281	&	0.951	&	0.534	&	1.936	&	1.754	&	W	&	15	\\	
	GZ~And		&	0.3050	&	-0.5157	&	1.115	&	0.593	&	0.532	&	0.990	&	0.742	&	2.278	&	4.616	&	W	&	2	\\	
	FU~Dra			&	0.3067	&	-0.5133	&	1.178	&	0.312	&	0.265	&	1.085	&	0.594	&	2.185	&	2.691	&	W	&	5	\\	
	TW~Cet		&	0.3117	&	-0.5063	&	1.284	&	0.680	&	0.530	&	1.053	&	0.788	&	2.422	&	5.861	&	W	&	16	\\	
	TY~Boo		&	0.3171	&	-0.4988	&	0.930	&	0.400	&	0.430	&	0.975	&	0.664	&	2.151	&	2.860	&	W	&	17	\\	
	SW~Lac		&	0.3207	&	-0.4939	&	1.240	&	0.964	&	0.777	&	1.028	&	0.917	&	2.565	&	7.795	&	W	&	4	\\	
	YY~Eri			&	0.3215	&	-0.4928	&	1.540	&	0.620	&	0.403	&	1.172	&	0.775	&	2.552	&	6.274	&	W	&	18,19,20\\	
	AO~Cam		&	0.3299	&	-0.4816	&	1.119	&	0.486	&	0.434	&	1.064	&	0.728	&	2.351	&	3.979	&	W	&	2	\\	
	AB~And			&	0.3319	&	-0.4790	&	1.042	&	0.595	&	0.571	&	1.018	&	0.788	&	2.377	&	4.516	&	W	&	2	\\	
	W~UMa		&	0.3336	&	-0.4768	&	1.190	&	0.570	&	0.479	&	1.084	&	0.775	&	2.443	&	4.831	&	W	&	7,21	\\	
	RZ~Com		&	0.3385	&	-0.4704	&	1.108	&	0.484	&	0.437	&	1.078	&	0.739	&	2.386	&	3.968	&	W	&	7	\\	
	GM~Dra		&	0.3387	&	-0.4702	&	1.213	&	0.219	&	0.181	&	1.220	&	0.564	&	2.304	&	2.037	&	W	&	4	\\	
	VW~Boo		&	0.3422	&	-0.4657	&	0.980	&	0.420	&	0.429	&	1.045	&	0.710	&	2.302	&	3.191	&	W	&	17	\\	
	V757~Cen		&	0.3432	&	-0.4645	&	1.000	&	0.690	&	0.690	&	1.010	&	0.853	&	2.456	&	5.028	&	W	&	22,23	\\	
	V781~Tau		&	0.3449	&	-0.4623	&	1.237	&	0.501	&	0.405	&	1.141	&	0.756	&	2.487	&	4.482	&	W	&	24	\\	
	ET~Leo			&	0.3465	&	-0.4603	&	1.586	&	0.542	&	0.342	&	1.265	&	0.776	&	2.669	&	5.820	&	W	&	6	\\	
	BV~Dra			&	0.3501	&	-0.4558	&	0.997	&	0.401	&	0.402	&	1.073	&	0.709	&	2.336	&	3.124	&	W	&	15	\\	
	AC~Boo		&	0.3524	&	-0.4530	&	1.403	&	0.565	&	0.403	&	1.208	&	0.798	&	2.630	&	5.540	&	W	&	25	\\	
	QW~Gem		&	0.3581	&	-0.4460	&	1.314	&	0.438	&	0.333	&	1.217	&	0.739	&	2.557	&	4.203	&	W	&	1	\\	
	V829~Her		&	0.3582	&	-0.4459	&	0.856	&	0.372	&	0.435	&	1.028	&	0.703	&	2.272	&	2.618	&	W	&	3	\\	
	AH~Cnc			&	0.3604	&	-0.4432	&	1.290	&	0.540	&	0.419	&	1.188	&	0.799	&	2.606	&	5.025	&	W	&	22	\\	
	BB~Peg		&	0.3615	&	-0.4419	&	1.424	&	0.550	&	0.386	&	1.240	&	0.804	&	2.678	&	5.514	&	W	&	5	\\	
	AE~Phe			&	0.3624	&	-0.4408	&	1.366	&	0.629	&	0.460	&	1.204	&	0.846	&	2.692	&	6.033	&	W	&	7,26,27\\	
	LS~Del			&	0.3638	&	-0.4391	&	1.068	&	0.399	&	0.374	&	1.135	&	0.725	&	2.436	&	3.319	&	W	&	28,29	\\	
	AM~Leo		&	0.3658	&	-0.4368	&	1.386	&	0.623	&	0.449	&	1.220	&	0.848	&	2.715	&	6.068	&	W	&	30,31	\\	
	V752~Cen		&	0.3702	&	-0.4316	&	1.320	&	0.410	&	0.311	&	1.256	&	0.738	&	2.604	&	4.013	&	W	&	32	\\	
	U~Peg			&	0.3748	&	-0.4262	&	1.149	&	0.379	&	0.330	&	1.201	&	0.726	&	2.519	&	3.380	&	W	&	33	\\	
	EE~Cet			&	0.3799	&	-0.4203	&	1.391	&	0.438	&	0.315	&	1.298	&	0.768	&	2.698	&	4.474	&	W	&	34	\\	
	TX~Cnc			&	0.3829	&	-0.4169	&	0.792	&	0.420	&	0.530	&	1.028	&	0.770	&	2.365	&	2.809	&	W	&	7	\\	
	BH~Cas		&	0.4059	&	-0.3916	&	0.743	&	0.352	&	0.474	&	1.057	&	0.752	&	2.377	&	2.329	&	W	&	35,36	\\	
	SS~Ari			&	0.4060	&	-0.3915	&	1.343	&	0.406	&	0.302	&	1.347	&	0.783	&	2.779	&	4.155	&	W	&	37	\\	
	AH~Vir			&	0.4075	&	-0.3899	&	1.360	&	0.412	&	0.303	&	1.356	&	0.788	&	2.798	&	4.256	&	W	&	38	\\	
	HT~Vir			&	0.4077	&	-0.3897	&	1.284	&	1.046	&	0.815	&	1.217	&	1.108	&	3.066	&	9.314	&	W	&	5	\\	
	UV~Lyn			&	0.4150	&	-0.3820	&	1.344	&	0.501	&	0.373	&	1.338	&	0.854	&	2.870	&	5.077	&	W	&	5	\\	
	V2357~Oph		&	0.4156	&	-0.3813	&	1.191	&	0.288	&	0.242	&	1.346	&	0.708	&	2.669	&	2.786	&	W	&	6	\\	
	V842~Her		&	0.4190	&	-0.3778	&	1.360	&	0.353	&	0.260	&	1.404	&	0.762	&	2.818	&	3.722	&	W	&	39	\\	
	ER~Ori			&	0.4234	&	-0.3732	&	1.385	&	0.765	&	0.552	&	1.320	&	1.007	&	3.061	&	7.643	&	W	&	40	\\	
	EF~Boo		&	0.4295	&	-0.3670	&	1.547	&	0.792	&	0.512	&	1.392	&	1.026	&	3.179	&	8.634	&	W	&	4	\\	
	VY~Sex		&	0.4434	&	-0.3532	&	1.423	&	0.449	&	0.316	&	1.450	&	0.859	&	3.015	&	4.901	&	W	&	6	\\	
	EZ~Hya		&	0.4498	&	-0.3470	&	1.370	&	0.350	&	0.255	&	1.478	&	0.796	&	2.959	&	3.802	&	W	&	41	\\	
	V502~Oph		&	0.4534	&	-0.3435	&	1.297	&	0.481	&	0.371	&	1.403	&	0.894	&	3.008	&	4.905	&	W	&	7	\\	
	AA~UMa		&	0.4680	&	-0.3298	&	1.419	&	0.773	&	0.545	&	1.424	&	1.079	&	3.294	&	8.128	&	W	&	32	\\	
	V728~Her		&	0.4713	&	-0.3267	&	1.654	&	0.295	&	0.178	&	1.688	&	0.776	&	3.182	&	3.769	&	W	&	42	\\	
	DN~Cam		&	0.4983	&	-0.3025	&	1.849	&	0.818	&	0.442	&	1.653	&	1.140	&	3.667	&	10.720	&	W	&	2	\\	
\hline
\hline
\end{tabular}
\end{small}   
\end{flushleft}
\begin{small}   
1: Kreiner et al. 2003, 2: Baran et al. 2004, 3: Zola et al. 2004, 4: Gazeas et al. 2005, 5: Zola et al. 2005, 6: Gazeas et al. 2006, 7: Hilditch et al. 1988, 8: Zhang et al. 2004, 9: Milone et al. 1987, 
10: Maceroni et al. 1981, 11: Khajavi et al. 2002, 12: Samec \& Hube1991, \\
13: Yakut et al. 2003, 14: Hilditch et al. 1992, 15: Kaluzny \& Rucinski 1986, 16: Russo et al. 1982, 17: Rainger et al. 1990a, 18: Nesci et al. 1986, 19: Yang \& Liu 1999, 20: Maceroni et al. 1994, 
21: Rucinski et al. 1993, 22: Maceroni et al. 1984, 23: Kaluzny 1984, 24: Yakut et al. 2005, 25: Mancuso et al. 1978, 26: Barnes et al. 2004, 27: Duerbeck 1978, 28: Lu \& Rucinski 1999, 29: Sezer et al. 1985, \\
30: Binnendijk 1984, 31: Hrivnak 1993, 32: Barone et al. 1993, 33: Pribulla \& Vanko 2002, 34: Rucinski et al. 2002, 35: Zola et al. 2001, 36: Metcalfe 1999, 37: Kim et al. 2003, 38: Lu \& Rucinski 1993, 
39: Rucinski \& Lu 1999, 40: Goecking et al. 1994, 41: Yang \& Qian 2004, 42: Nelson et al. 1995, 43: Rainger et al. 1990b, 44: Qian \& Yang 2005, 45: Lapasset \& Gomez 1990, 46: McLean \& Hilditch 1983, 
47: Gazeas et al. 2007, 48: Zola et al. (under prep.), 49: Bilir et al. 2005, 50: Rucinski et al., 2003, 51: Maceroni et al. 1996, 52: Lu et al. 2001, 53: Pribulla et al., 2002, 54: Niarchos \& Manimanis 2003, 
55: Milone et al. 1995, 56: Yang \& Liu 2003, 57: Yang \& Liu 2003b, \\
58: Rovithis-Livaniou et al. 2001, 59: Pribulla et al. 2001, 60: Pribulla et al. 1999, 61: Hilditch et al. 1989, (continued below Table 2)
\end{small}   
\end{table*}

\begin{table*}
\caption{Results derived from the light curve modeling for 60 A-type contact binaries}
\begin{flushleft}
\begin{small}   
\begin{tabular}{lccccccccccc}
\hline
Name			&$P_{\rm{rot}}$	&logP		&$M_{1}$		&$M_{2}$	&q		&$R_{1}$	&$R_{2}$	&a		&$H_{\rm{orb}}$		&Type	&Ref	\\	
			&(days)		&		&$(M_{\odot})$		&$(M_{\odot})$&		&$(R_{\odot})$	&$(R_{\odot})$	&$(R_{\odot})$	&($10^{-51}$ gcm$^{2}$s$^{-1}$)   &	&	\\	
\hline	
	OU~Ser			&	0.2968	&	-0.5275	&	1.109	&	0.192	&	0.173	&	1.089	&	0.494	&	2.043	&	1.613	&	A	&	5	\\	
	TZ~Boo		&	0.2972	&	-0.5270	&	0.783	&	0.104	&	0.133	&	0.999	&	0.404	&	1.800	&	0.701	&	A	&	7,43	\\	
	SX~Crv			&	0.3166	&	-0.4995	&	1.246	&	0.098	&	0.079	&	1.287	&	0.416	&	2.156	&	0.935	&	A	&	3	\\	
	FG~Hya		&	0.3278	&	-0.4844	&	1.414	&	0.157	&	0.111	&	1.325	&	0.495	&	2.325	&	1.633	&	A	&	44,28	\\	
	EQ~Tau		&	0.3413	&	-0.4668	&	1.233	&	0.551	&	0.447	&	1.121	&	0.777	&	2.492	&	4.854	&	A	&	5	\\	
	V508~Oph		&	0.3448	&	-0.4624	&	1.010	&	0.520	&	0.515	&	1.043	&	0.770	&	2.383	&	3.963	&	A	&	45	\\	
	GR~Vir			&	0.3470	&	-0.4597	&	1.376	&	0.168	&	0.122	&	1.350	&	0.526	&	2.401	&	1.743	&	A	&	4	\\	
	CK~Boo		&	0.3552	&	-0.4496	&	1.442	&	0.155	&	0.107	&	1.412	&	0.521	&	2.466	&	1.679	&	A	&	6	\\	
	VZ~Lib			&	0.3583	&	-0.4458	&	1.480	&	0.378	&	0.255	&	1.303	&	0.702	&	2.609	&	4.007	&	A	&	3	\\	
	DZ~Psc		&	0.3661	&	-0.4364	&	1.352	&	0.183	&	0.135	&	1.375	&	0.560	&	2.483	&	1.902	&	A	&	4	\\	
	V410~Aur		&	0.3664	&	-0.4361	&	1.270	&	0.173	&	0.136	&	1.346	&	0.550	&	2.434	&	1.725	&	A	&	6	\\	
	V417~Aql		&	0.3703	&	-0.4314	&	1.377	&	0.498	&	0.362	&	1.254	&	0.790	&	2.675	&	4.951	&	A	&	4	\\	
	XY~Boo		&	0.3706	&	-0.4311	&	0.934	&	0.147	&	0.157	&	1.205	&	0.524	&	2.227	&	1.191	&	A	&	46,77	\\	
	HV~Aqr			&	0.3745	&	-0.4265	&	1.366	&	0.198	&	0.145	&	1.390	&	0.584	&	2.537	&	2.082	&	A	&	47	\\	
	RT~LMi			&	0.3749	&	-0.4261	&	1.298	&	0.476	&	0.367	&	1.238	&	0.784	&	2.648	&	4.563	&	A	&	48	\\	
	YY~CrB		&	0.3766	&	-0.4241	&	1.393	&	0.339	&	0.243	&	1.327	&	0.700	&	2.634	&	3.521	&	A	&	4	\\	
	HX~UMa		&	0.3792	&	-0.4211	&	1.333	&	0.387	&	0.290	&	1.289	&	0.736	&	2.640	&	3.864	&	A	&	49,50	\\	
	HN~UMa		&	0.3825	&	-0.4173	&	1.279	&	0.179	&	0.140	&	1.385	&	0.572	&	2.513	&	1.817	&	A	&	5	\\	
	BI~CVn			&	0.3842	&	-0.4154	&	1.646	&	0.679	&	0.413	&	1.346	&	0.900	&	2.945	&	7.604	&	A	&	51	\\	
	AU~Ser			&	0.3865	&	-0.4129	&	0.921	&	0.646	&	0.701	&	1.063	&	0.904	&	2.592	&	4.626	&	A	&	51	\\	
	EX~Leo			&	0.4086	&	-0.3887	&	1.557	&	0.309	&	0.198	&	1.487	&	0.716	&	2.852	&	3.595	&	A	&	52,53	\\	
	V839~Oph		&	0.4090	&	-0.3883	&	1.572	&	0.462	&	0.294	&	1.431	&	0.821	&	2.937	&	5.275	&	A	&	6	\\	
	V566~Oph		&	0.4096	&	-0.3876	&	1.469	&	0.357	&	0.243	&	1.429	&	0.753	&	2.836	&	3.951	&	A	&	54	\\	
	QX~And		&	0.4118	&	-0.3853	&	1.176	&	0.236	&	0.201	&	1.360	&	0.658	&	2.612	&	2.282	&	A	&	55	\\	
	RZ~Tau		&	0.4157	&	-0.3812	&	1.634	&	0.882	&	0.540	&	1.380	&	1.042	&	3.187	&	9.805	&	A	&	56,57	\\	
	Y~Sex			&	0.4198	&	-0.3770	&	1.210	&	0.220	&	0.182	&	1.405	&	0.651	&	2.657	&	2.194	&	A	&	56	\\	
	AK~Her			&	0.4215	&	-0.3752	&	1.310	&	0.300	&	0.229	&	1.411	&	0.724	&	2.772	&	3.117	&	A	&	58	\\	
	EF~Dra			&	0.4240	&	-0.3726	&	1.813	&	0.289	&	0.159	&	1.642	&	0.719	&	3.041	&	3.810	&	A	&	59	\\	
	AP~Leo			&	0.4304	&	-0.3661	&	1.460	&	0.434	&	0.297	&	1.442	&	0.832	&	2.967	&	4.794	&	A	&	1	\\	
	AW~UMa		&	0.4387	&	-0.3578	&	1.280	&	0.090	&	0.070	&	1.633	&	0.503	&	2.697	&	0.977	&	A	&	60	\\	
	V776~Cas		&	0.4404	&	-0.3561	&	1.750	&	0.342	&	0.195	&	1.628	&	0.779	&	3.114	&	4.414	&	A	&	5	\\	
	UX~Eri			&	0.4453	&	-0.3513	&	1.430	&	0.534	&	0.373	&	1.431	&	0.914	&	3.072	&	5.773	&	A	&	4	\\	
	TV~Mus		&	0.4457	&	-0.3510	&	1.316	&	0.157	&	0.119	&	1.575	&	0.608	&	2.793	&	1.720	&	A	&	61	\\	
	DK~Cyg		&	0.4707	&	-0.3273	&	1.741	&	0.533	&	0.306	&	1.619	&	0.946	&	3.347	&	6.806	&	A	&	2	\\	
	AQ~Psc			&	0.4756	&	-0.3227	&	1.682	&	0.389	&	0.231	&	1.660	&	0.856	&	3.267	&	4.968	&	A	&	6	\\	
	NN~Vir			&	0.4807	&	-0.3181	&	1.730	&	0.850	&	0.491	&	1.563	&	1.131	&	3.540	&	10.413	&	A	&	4	\\	
	EL~Aqr			&	0.4814	&	-0.3175	&	1.563	&	0.317	&	0.203	&	1.657	&	0.806	&	3.189	&	3.901	&	A	&	62	\\	
	XZ~Leo			&	0.4877	&	-0.3118	&	1.742	&	0.586	&	0.336	&	1.642	&	1.001	&	3.454	&	7.517	&	A	&	6	\\	
	AH~Aur			&	0.4943	&	-0.3060	&	1.674	&	0.283	&	0.169	&	1.760	&	0.790	&	3.289	&	3.713	&	A	&	4	\\	
	OO~Aql			&	0.5068	&	-0.2952	&	1.040	&	0.880	&	0.846	&	1.308	&	1.212	&	3.323	&	7.279	&	A	&	63	\\	
	V401~Cyg		&	0.5827	&	-0.2346	&	1.679	&	0.487	&	0.290	&	1.854	&	1.057	&	3.797	&	6.544	&	A	&	34,64	\\	
	eps~Cra		&	0.5914	&	-0.2281	&	1.720	&	0.220	&	0.128	&	2.064	&	0.820	&	3.696	&	3.157	&	A	&	65	\\	
	V351~Peg		&	0.5933	&	-0.2267	&	2.327	&	0.838	&	0.360	&	2.046	&	1.286	&	4.361	&	13.837	&	A	&	66	\\	
	AQ~Tuc			&	0.5948	&	-0.2256	&	1.930	&	0.690	&	0.358	&	1.927	&	1.207	&	4.101	&	10.072	&	A	&	67	\\	
	V402~Aur		&	0.6035	&	-0.2193	&	1.638	&	0.327	&	0.200	&	1.960	&	0.947	&	3.763	&	4.481	&	A	&	3	\\	
	RR~Cen		&	0.6057	&	-0.2177	&	1.854	&	0.389	&	0.210	&	2.037	&	1.005	&	3.942	&	5.780	&	A	&	7,68	\\	
	UZ~Leo		&	0.6180	&	-0.2090	&	2.074	&	0.629	&	0.303	&	2.060	&	1.198	&	4.251	&	9.890	&	A	&	39,69	\\	
	V535~Ara		&	0.6293	&	-0.2011	&	1.520	&	0.460	&	0.303	&	1.880	&	1.093	&	3.879	&	5.916	&	A	&	70	\\	
	BD~+14~5016		&	0.6369	&	-0.1959	&	1.531	&	0.387	&	0.253	&	1.936	&	1.038	&	3.869	&	5.087	&	A	&	71	\\	
	FP~Boo		&	0.6404	&	-0.1935	&	1.614	&	0.154	&	0.095	&	2.199	&	0.771	&	3.779	&	2.197	&	A	&	6	\\	
	AG~Vir			&	0.6427	&	-0.1920	&	1.610	&	0.510	&	0.317	&	1.934	&	1.147	&	4.024	&	6.839	&	A	&	72	\\	
	S~Ant			&	0.6483	&	-0.1882	&	1.940	&	0.760	&	0.392	&	2.026	&	1.322	&	4.388	&	11.362	&	A	&	51	\\	
	FN~Cam		&	0.6771	&	-0.1693	&	2.402	&	0.532	&	0.221	&	2.377	&	1.202	&	4.643	&	9.718	&	A	&	62,73	\\	
	HV~UMa		&	0.7107	&	-0.1483	&	2.800	&	0.500	&	0.179	&	2.646	&	1.217	&	4.988	&	10.404	&	A	&	74	\\	
	V592~Per		&	0.7157	&	-0.1453	&	1.743	&	0.678	&	0.389	&	2.090	&	1.360	&	4.519	&	9.760	&	A	&	5	\\	
	V1073~Cyg		&	0.7859	&	-0.1046	&	1.600	&	0.510	&	0.319	&	2.206	&	1.312	&	4.595	&	7.279	&	A	&	75	\\	
	V2388~Oph		&	0.8023	&	-0.0957	&	1.648	&	0.306	&	0.186	&	2.394	&	1.120	&	4.541	&	4.647	&	A	&	76	\\	
	TY~Pup		&	0.8192	&	-0.0866	&	2.200	&	0.720	&	0.327	&	2.515	&	1.514	&	5.264	&	12.856	&	A	&	51	\\	
	II~UMa			&	0.8250	&	-0.0835	&	2.238	&	0.385	&	0.172	&	2.723	&	1.232	&	5.103	&	7.265	&	A	&	62	\\	
	V921~Her		&	0.8774	&	-0.0568	&	2.068	&	0.505	&	0.244	&	2.660	&	1.405	&	5.283	&	9.046	&	A	&	6	\\	
\hline
\hline
\end{tabular}
\end{small}   
\end{flushleft}
\begin{small}   
62: Rucinski et al. 2001, 63: Hrivnak et al. 2001, 64: Wolf et al. 2000, 65: Goecking \& Duerbeck 1993, 66: Gazeas et al. (under prep.), \\
67: Hilditch 1986, 68: King \& Hilditch 1984, 69: Vink\'o et al. 1996, 70: Leung et al. 1978, 71: Maciejewski et al. 2003, 72: Bell et al. 1990, \\
73: Vanko et al. 2001, 74: Cs\'ak et al. 2000, 75: Ahn et al. 1992, 76: Yakut et al. 2004, 77: Awadalla \&Yamasaki 1984, 78:Awadalla 1989, \\
79:Binnendijk 1963
\end{small}
\end{table*}
%-------------------End of Table 1 ------------------------------------

\begin{figure}
\begin{center}
\rotatebox{0}{\scalebox{0.45}{\includegraphics{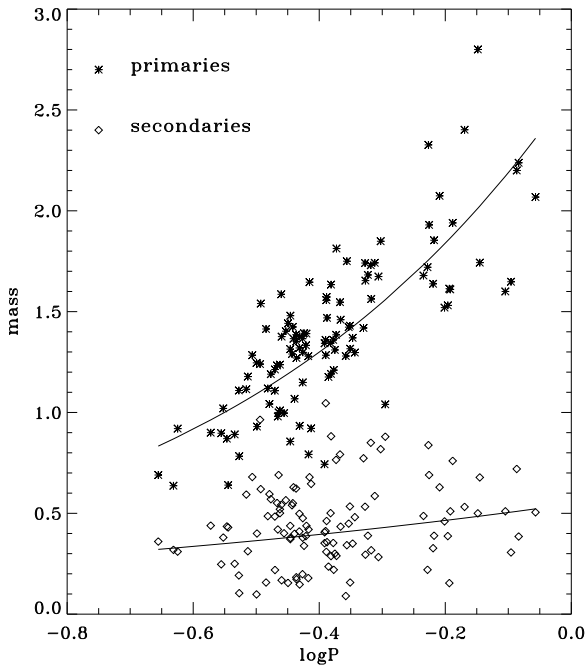}}}
\caption{\label{fig1}
The mass distribution of the components of all 112 contact binaries in
our sample. The masses of the primary components are plotted with asterisks,
while those of secondaries with diamonds. Solid lines are least square
fits given by Eqs.~(4)-(5).}
\end{center}
\end{figure} 

\citet{Lucy1976} assumed in his model of a cool contact binary that both
components of W-type stars are located on ZAMS and they evolve via Thermal
Relaxation Oscillations (TRO) with a secular mass transfer from the
secondary to primary component until the primary reaches a limiting mass
for CNO cycle to dominate hydrogen burning process. The primary evolves
then off the ZAMS and increases its radius so that both components can fill
their critical Roche lobes being in thermal equilibrium.  A similar
conclusion was reached by \citet{Hilditch1988} and \citet{Hilditch1989}. A
different view was taken by \citet{Maceroni1996} who analyzed properties of
a numerous sample of W~UMa stars of both types and concluded ``... that
most A-type systems have no evolutionary link with the W-types, as they
have too large total mass and/or angular momentum to be the result of
evolution of W-types towards smaller mass ratios.'' A recent discussion of
over one hundred cool contact binaries with accurately determined
parameters, carried out by \citet{GN2006}, confirmed the conclusion of
\citet{Maceroni1996}. If W-type stars cannot evolve into A-type, can the
opposite be true? W~UMa stars are magnetically very active and it is
generally accepted that they lose mass and AM via magnetized wind 
\citep{Stepien1995, Stepien2006a, Stepien2006b, YE2005}. 
Is it possible that A-type stars originate from short-period detached binaries and, after losing a
fraction of mass and AM they evolve into W-type systems? Or, perhaps,
both types evolve in their own ways, i. e. remaining A-type or W-type from
the formation of contact configuration to a probable merging of both
components into a single, rapidly rotating star? 

Tables~1 and 2 list names, geometrical and physical data of 112 cool contact
binaries with accurately determined parameters. It is known that the
parameters of W~UMa-type binaries fulfill several relations. Some of them
result from the fact that they are contact systems and their primaries are
MS stars (see below) but others are not obvious \`a priori. Instead, they
result from the correlations shown by the observations. We will discuss
them in turn.

\begin{figure}
\begin{center}
\rotatebox{0}{\scalebox{0.45}{\includegraphics{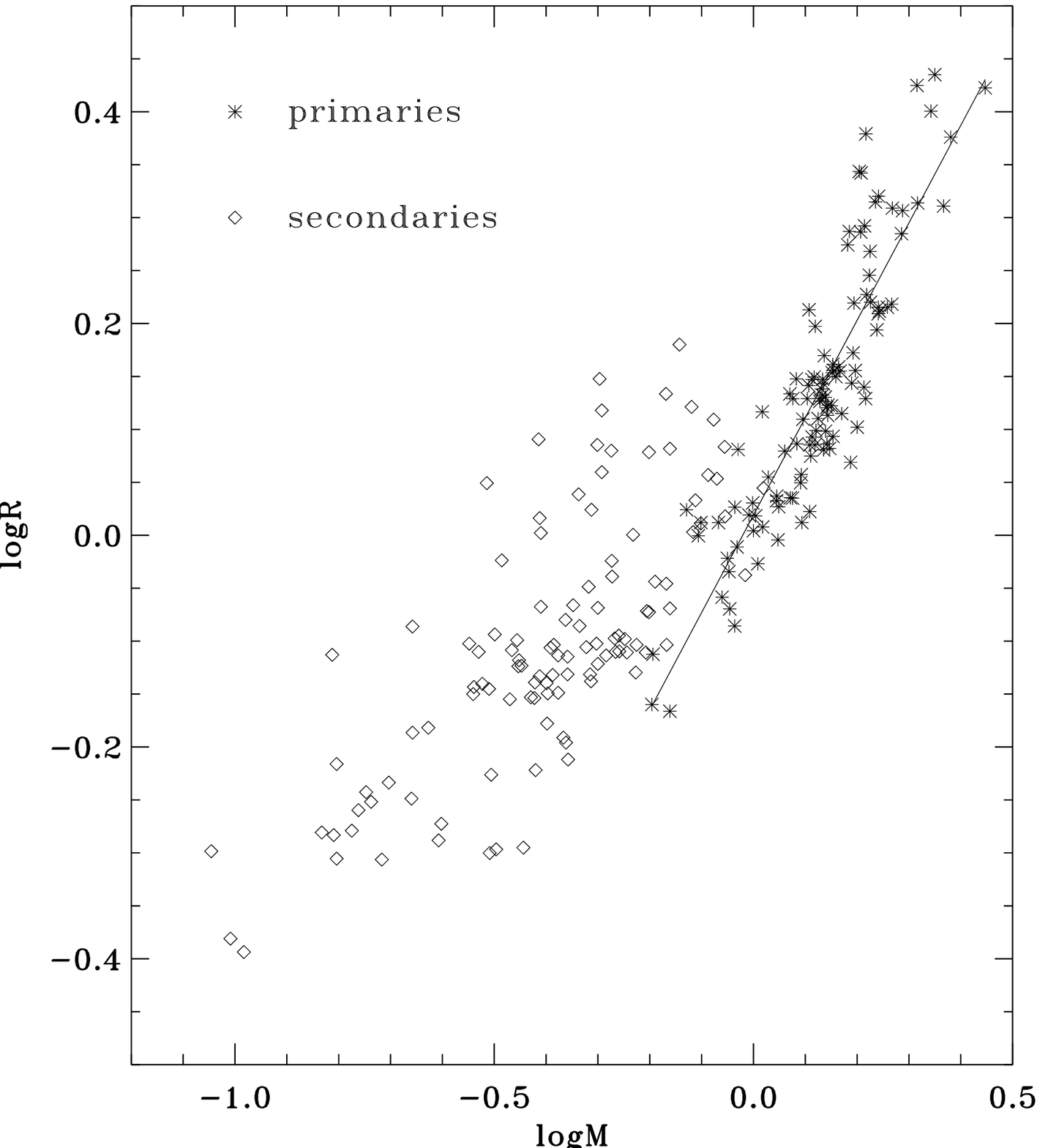}}}
\caption{\label{fig2}
The radius distribution of the components of contact
binaries versus mass. Symbols are as in Fig.~1. Straight line gives the
least square fitted  
mass-radius relation for primary components.}
\end{center}
\end{figure} 
%----------------------- Fig. 2 end ---------------------------------

An important period-color relation was discovered by \citet{Eggen1961}. Its
recent version reads \citep{Wang1994}

\begin{equation}
(B-V)_0 = 0.062 - 1.31\log P\,,
\end{equation}

where orbital period  $P$ is in days. \citet{RD1997} derived a calibration
of absolute magnitude of W~UMa-type stars in terms of color and period

\begin{equation}
M_V = -4.44\log P + 3.02(B-V)_0 + 0.12\,.
\end{equation}

Combining both equations we obtain

\begin{equation}
M_V = -8.4\log P + 0.31\,.
\end{equation}
 
Eq.(3) shows that the knowledge of the orbital period is sufficient to
calculate the absolute magnitude of a W~UMa-type binary with a fair
accuracy. A similar, but steeper relation was recently derived by
\citet{Ruc2006} for stars with $\log P < -0.25$.
\citet{GN2006} noticed already that mass of the primary components of W
UMa-type binaries increases steeply with increasing period, whereas mass
of secondaries is nearly period independent and varies between 0 and 1
$M_{\odot}$.  Fig.~1 shows the period-mass relations for both components
of the systems from Tables~1 and 2. Power law relations can be fitted to the
data 

\begin{equation}
\log M_{\rmn{1}} = (0.755\pm 0.059)\log P + (0.416\pm 0.024)\,,
\end{equation}

\begin{equation}
\log M_{\rmn{2}} = (0.352\pm 0.166)\log P - (0.262\pm 0.067)\,.
\end{equation}

As we see, the knowledge of the orbital period suffices to estimate not
only the absolute magnitude of the system but also
masses of both components with a reasonable accuracy of about 15\%. The
above relations may be useful
when statistically analyzing data from mass photometry programs, such as
ASAS, OGLE or MACHO. A-type binaries follow the same relations as
W-type.

Several authors noticed that primaries of cool contact binaries obey
mass-radius relation for MS stars. Our data confirm this result. Fig.~2 is
a plot of radii of both components of binaries from Tables~1 and 2 versus mass. A
power law fit to the primary radii gives the relation $R_{\rmn{1}} \propto
M_{\rmn{1}}^\alpha$ where $\alpha = 0.92\pm 0.04$. This is very close to
the exponent $\alpha =0.977$ of the empirical mass-radius relation for
single MS stars with masses lower than 1.8 $M_{\odot}$
\citep{GZ1985}. Secondaries are substantially oversized and do not follow
any simple mass-radius relation.

Orbital parameters of W~UMa-type contact binaries obey some basic
relations resulting from the Roche model. These are: the third Kepler law

\begin{equation}
P = 0.1159a^{3/2}M^{1/2}\,,
\end{equation}

where $M = M_{\rmn{1}} + M_{\rmn{2}}$, the total mass, and $a$, semiaxis, are
in solar units and $P$ in days, the expression for orbital angular momentum

\begin{equation}
H_{\rm{orb}} = 1.24\times 10^{52}M^{5/3}P^{1/3}q(1+q)^{-2}\,,
\end{equation}

with $H_{\rm{orb}}$ in cgs units, and finally the expressions for
critical Roche lobe sizes, approximated by \citet{Eggleton1983}, and
assumed to be identical with stellar radii

\begin{equation}
\frac{R_{\rmn{1}}}{a} = \frac{0.49q^{2/3}}{0.6q^{2/3}+\ln{(1 + q^{1/3})}}\,,
\end{equation}

\begin{equation}
\frac{R_{\rmn{2}}}{a} = \frac{0.49q^{-2/3}}{0.6q^{-2/3}+\ln{(1 +
q^{-1/3})}}\,.
\end{equation}

Using the empirical period-mass relations given by Eqs.~(4)-(5) and the
above formulas, the period-radius relations for primary and secondary
components can be numerically calculated. Fig.~3 shows these relations as
solid lines with the observed values of the component radii
over-plotted. The agreement between the computed relations and observed data
is perfect.
Two more approximate relations can also be derived. Neglecting
variability of right hand sides of Eqs.~(8)-(9) on $q$ (e.g. by
putting $q \equiv \overline q = 0.34$) we have: $H_{\rm{orb}} \propto
a^{1/2}M_1^{3/2}, P \propto a^{3/2}M_1^{-1/2}$ and $R_{\rmn{1}} \propto a$,
where we adopted $M = 1.34M_{\rmn{1}}$. Using the empirical relation
$R_{\rmn{1}} \propto M_1^\alpha$ we obtain for $\alpha \approx 1$

\begin{equation}
P \propto
M_1^{3\alpha/2} M_1^{-1/2} \propto M_1^{\frac{3\alpha - 1}{2}} \approx M_1\,,
\end{equation}

\begin{equation}
H_{\rm{orb}} \propto M_1^{\alpha/2}M_1^{3/2}
\propto M_1^{\frac{\alpha +3}{2}} \approx M_1^2\,. 
\end{equation}

where the final exponents are rounded to the nearest integer.  We finally
obtain $H_{\rm{orb}} \propto P^2$ and $M \propto P$. The plot of the 
total observed mass versus period, given by
\citet{GN2006} (see their Fig.~3) shows indeed the correlation in the
predicted sense. Note that the total masses of several near-contact
binaries, also plotted in that figure, are lower than those of genuine
contact binaries with the same orbital period. This is a consequence of the
fact that masses of primaries of near contact binaries are too low, hence
their radii are too small to fill up their critical Roche lobes (assuming
the primaries are MS stars). It is interesting to see that the
$P-H_{\rm{orb}}$ plot of contact binaries, given in Fig.~4, shows much
poorer correlation as also shown by \citet{GN2006} (see their Fig.~4).
Values of AM for individual stars fill a part of the figure below diagonal
running from the lower left (short periods, low AM) to the upper right
corner (long periods, high AM). The average value of AM increases with
increasing period, as predicted, but the scatter increases as well. The
increasing scatter can be explained by a more sensitive dependence of AM on
mass ratio. The upper bound given by the diagonal corresponds to
binaries with maximum component masses for a given period. On the other hand,
binaries with extremely low secondary masses (hence mass ratios of
$q \approx$ 0.1) lie low in the figure. A-type binaries
listed in Table~2 cover a broader range of values of AM (from 0.7 to 13.8
$\times 10^{51}$ in cgs units) than W-type binaries (from 1.7 to 10.7$\times 10^{51}$ in
cgs units). This indicates that a picture of contact binary
evolution from A-type to W-type, or vice versa is too simplistic (see
\citealt{GN2006, Eker2006}).

Considering evolution of contact binaries one should note that an isolated
system can either preserve mass and AM (conservative evolution) or lose
{\em both} quantities simultaneously. Evolution can never move a binary
towards higher total mass and/or AM, as stressed by \citet{GN2006} but the
opposite direction is physically possible and in fact very likely. W~UMa
type binaries are very active so we expect strong magnetized winds carrying
away mass and AM. Old contact binaries should have lower total mass and AM
than the newly formed ones. The questions to answer are: where newly formed
contact binaries appear in period-AM and period-mass diagrams, and what do
their evolutionary tracks look like. In the next section we present an
evolutionary model of a contact binary which answers these questions.

%----------------------- Fig. 3 start --------------------------------
\begin{figure}
\begin{center}
\rotatebox{0}{\scalebox{0.45}{\includegraphics{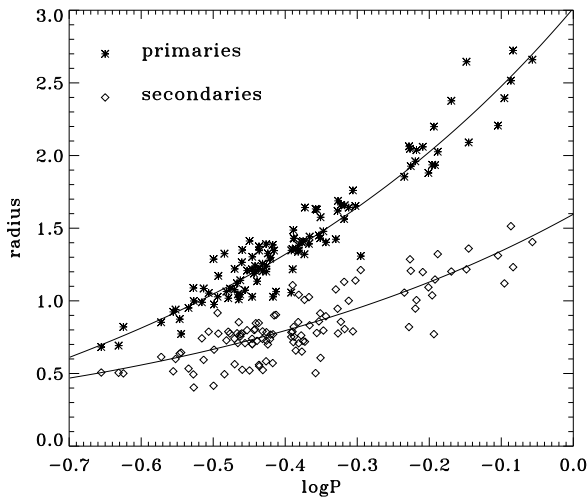}}}
\caption{\label{fig3}
The radius distribution of 112 contact binaries, symbols as in
Fig.~1. Solid lines are theoretically predicted  period-radius (see text).}
\end{center}
\end{figure} 
%----------------------- Fig. 3 end ---------------------------------

\section{Evolutionary model of a cool contact binary}

The present model is based on evolutionary scenario developed by
\citet{Stepien2004, Stepien2006a, Stepien2006b}. The scenario assumes that
W~UMa type stars originate from cool detached binaries with initial orbital
periods of a few days. If such binaries are formed in the process of
fragmentation, their minimum ZAMS periods are expected to be close to 2
days \citep{Stepien1995}. We consider only binaries in which both
components possess subphotospheric convection zones and rotate
synchronously. Such stars are very active and
drive magnetized winds. Any possible
proximity effects on the winds are neglected (this also holds when a binary
enters the contact phase). Synchronous rotation demands that AM lost by the
winds is ultimately drawn from orbital AM.

Neglecting spin AM of both components, compared to orbital AM (see Fig.~4
in \citealt{GN2006}), the AM loss (AML) rate of a close binary is given by
Eq.~(15) in \citet{Stepien2006b}\footnote{unnecessary factor $\omega$
appeared in that equation - the correct form is given in the present paper
and in astro-ph/0701529}

\begin{equation}
\frac{{\rm d}H_{\rm{orb}}}{{\rm d}t} = -4.9\times 10^{41}(R_1^2M_1 +R_2^2M_2)/P\,.
\end{equation}

Here AM is in cgs units, period in days, masses and radii in solar units
and time in years.  The formula is based on semi-empirically determined AML
rate of single, cool stars. The uncertainty of the numerical
coefficient is about 30 \%. Similarly as in that paper, the supersaturation
effect is allowed for by assuming $P \equiv$ 0.4 days in Eq.~(12) for
periods shorter than 0.4 days.

The adopted mass loss rate of each
component is based on empirical determination by \citet{Wood2002}

\begin{equation}
\dot M_{1,2} = -10^{-11}R_{1,2}^2\,,
\end{equation}

where mass loss rates are in $M_{\odot}$/year and radii in solar units. The
uncertainty of the numerical factor is of the order of two. A more detailed
discussion of this formula is given by
\citet{Stepien2006b}. \citet{Wood2005} 
announced recently the observations of stellar wind from $\xi$
Boo - a very active star, which indicate that the mass loss rate from the
most active stars may actually be lower than resulting from the above
formula. 

Much higher mass and AML rates are suggested by \citet{Demircan2006}. The
  rates are based on a controversial assumption that chromospherically
  active binaries are born with mean total masses of $\sim 4 M_{\odot}$ and
  mean orbital periods of $\sim$ 50 days. During the subsequent evolution
  the binaries lose mass and AM at high rates so that, at the age of 9 Gyr,
  their mean masses and periods decrease to $\sim 1.2 M_{\odot}$ and $\sim$
  1.6 days, respectively (see Fig.~2 in \citealt{Demircan2006}). We think
  that the rates derived by \citet{Demircan2006} are too high and weakly
  justified. Additional, convincing arguments in their favor are needed
  before one accepts them.

To simplify calculations we adopt in both formulas the parametric
approximation $R_s = M_s$, where $R_s$ and $M_s$ are stellar radius and
mass in solar units. Observations of low mass MS stars show that this
equality is satisfied to a good approximation \citep{LMR2005}. 
Evolutionary models show that $R_s$ of a star with a mass $\leq 1
M_{\odot}$ varies between about $0.9M_s$ at ZAMS to slightly less than
$1.3M_s$ at TAMS, with a time-weighted average close to $M_s$
\citep{Schaller1992}, so the applied approximation is also in 
a good agreement with
theoretical models. The parametric approximation was used only in these two
equations. The actual (time-dependent) values of stellar radii of both
components were calculated as a part of the evolutionary model of a
binary. They were interpolated from evolutionary models of single stars and
compared at each time step to the critical Roche lobes (see below).

The above equations were combined with equations describing the orbital binary
parameters (see the previous section) and integrated in time to follow the
evolution of binaries with various initial masses. The value of 2 d
was always adopted for an initial orbital period. The results indicate that
time needed for a cool, close binary with such a period
and the initial mass of the primary around 1.2-1.3 $M_{\odot}$ to
reach contact amounts to several Gyr i. e. it is close to the life time
of the primary on MS (see also \citealt{Stepien1995, Stepien2006a}).
Moreover, both time scales show similar mass dependence.  From Eq.~(12) 
the AML time
scale $\tau_{\rm{AML}} \propto M_1^{-3}$, if we ignore the dependence of
the time scale on parameters of a secondary. On the other hand, the MS life
time, resulting from the models by \citet{Schaller1992}, can be
approximated to within 5 \% by: $\tau_{\rm{ev}} =9.84M^{-\gamma}$, 
where $\gamma \approx 3$
for $1 \leq M \leq 1.3 M_{\odot}$ and $\gamma \approx 4$ for $0.8 \leq M <
1 M_{\odot}$.  A similar mass
dependence of both time scales means that their approximate equality holds
down to the least massive stars with MS life time shorter than the Hubble
time, i. e. about 0.9 $M_{\odot}$. In consequence, the
primary is close to, or even beyond TAMS at
the time when its critical Roche lobe reaches the stellar radius. 

\begin{table}
\centering
\caption{Results of model calculations}
\begin{tabular}{@{}lcclc@{}}
%\hline
%\footnote{Remark: (2) - mass transfer rate in the contact phase 
%increased by 7 \% compared to (1).}\\
ev. stage & age & masses & period  & orb. AM\\
& Gyr & $M_{\odot}$ & days & $\times 10^{51}$\\
\hline
initial & 0 & 1.3+1.1 & 2 & 16.7\\
start contact & 4.8 & 0.85+1.45 & 0.73 & 10.5\\
coalescence(1) & 6.0 & 0.43+1.85 & 0.54 & 6.2\\
\hline
initial & 0 & 1.3+1.1 & 2 & 16.7\\
start contact & 4.8 & 0.85+1.45 & 0.73 & 10.5\\
coalescence(2) & 6.5 & 0.22+2.07 & 0.76 & 3.8\\
\hline
initial & 0 & 1.3+0.7 & 2 & 11.3\\
start contact & 4.6 & 0.66+1.27 & 0.49 & 6.5\\
coalescence & 5.6 & 0.23+1.69 & 0.61 & 3.3\\
\hline
initial & 0 & 1.0+0.8 & 2 & 10.3\\
start contact & 8.7 & 0.78+0.89 & 0.38 & 5.2\\
coalescence & 10.3 & 0.15+1.50 & 0.27 & 1.5\\
\hline
initial & 0 & 1.0+0.5 & 2 & 6.8\\
start contact & 8.8 & 0.49+0.86 & 0.28 & 3.1\\
coalescence & 10.1 & 0.14+1.20 & 0.31 & 1.3\\
\hline
initial & 0 & 0.9+0.45 & 2 & 5.7\\
start contact & 12.5 & 0.57+0.67 & 0.26 & 2.8\\
coalescence & 13.0 & 0.49+0.74 & 0.23 & 0.7\\
\hline
\end{tabular}
Remark: (2) - mass transfer rate in the contact phase
increased by 7 \% compared to (1).
\end{table}

Following the Roche lobe overflow (RLOF) mass transfer to the secondary
component begins. Whatever are the details of this process, our model
assumes that it ends when both stars reach thermal equilibrium with radii
not exceeding their Roche lobes, similarly as in case of Algols. We assume
that mass is transferred on a thermal time scale, i.e. during $\sim 10^8$
years, except for the least massive binary for which shorter time scale of
$\sim 10^7$ years was adopted.  Equilibrium radii of both stars become
smaller than sizes of the Roche lobes only after mass ratio reversal. It is
likely that additional AM and mass loss occur when $q \approx 1$ and the
semiaxis is at its local minimum but the accurate modelling of the common
envelope phase is still beyond the present capabilities \citep{YE2005}. 
To avoid introducing additional arbitrary parameters describing the
possible mass and AM loss during the common envelope phase we assumed
conservative mass exchange (apart from the stellar winds). However, in two
most massive cases modeled, the binaries emerged as near-contact binaries
of an Algol type after the mass exchange phase of $10^8$ years. The systems
contained too much AM to form contact binaries. After some additional time
such short period Algols should lose enough AM to turn into a contact
configuration \citep{Stepien2006a}. To skip this semi-detached phase, the
mass exchange phase was artificially extended in time by lowering the mass
transfer rate so that the right amount of AM was lost via winds and a
contact system was formed as a post-mass-exchange equilibrium
configuration. The extended mass exchange phase took $4-6\times 10^8$
years. This manipulation was not needed for the three less massive binaries
which emerged from the mass exchange phase as contact binaries. A newly
formed contact binary consists of the present secondary (originally more
massive) with a radius equal to its TAMS value, and the present primary
(originally less massive) with a radius equal to its ZAMS value. The
adopted mass transfer rates lie in the interval $1 - 5\times 10^{-9}
M_{\odot}$/year.

The last evolutionary stage considered by us is the binary evolution in
contact. Mass and AM loss are governed by Eqs.~(12)-(13), as before, but
mass is also transferred from the present secondary (hydrogen depleted) to
the present primary, virtually unevolved after gaining the hydrogen rich
matter during the fast mass exchange phase. The mass transfer from the
secondary is caused by its evolutionary expansion and it ultimately leads
to $q \rightarrow 0$. In the lack of precise evolutionary models of a
contact binary, the mass transfer rate in the contact phase is treated as a
free parameter, similarly as in the previous phase. As it turned out, its
correct value required a very fine tuning. Too high rate results in a rapid
increase of the orbital period and the corresponding increase of both
critical Roche lobes. Even if the present secondary could expand fast
enough to fill its rapidly increasing Roche lobe, the present primary
cannot keep pace with its growing Roche lobe (remember that both components
are assumed to be in thermal equilibrium), and the binary would transform
into an ordinary Algol with a period exceeding 1 day. On the other hand, too
low mass transfer rate in the presence of a fixed AML results in a
period decrease, and the overflow of the critical Roche surface.  Different
estimates indicate that a contact configuration exists for one or a few Gyr
\citep{Mochnacki1981, Bilir2005, Stepien2006a}. To keep a contact
configuration for such a time a value of the mass transfer rate must
be individually adjusted with a relatively high precision. The resulting
values lie in the interval $3-4\times 10^{-10} M_{\odot}$/year, i. e. they
are about ten times lower than in the fast mass transfer phase. During this
phase, the radius of the secondary was kept equal to the size of its Roche
lobe, whereas the radius of the primary was assumed to increase a little
due to evolutionary effects (specifically, from $R_{\rm{ZAMS}} = 0.9M$ to
$R = M$).  The computations were stopped when the Roche lobe of the primary
became smaller than its radius and the second RLOF occurred.  In most cases
this takes place when the mass
ratio approaches a critical value of 0.1 beyond which coalescence of both
components is expected \citep{Rasio1995} but two models behaved differently (see
below).

\section{Discussion}

The model computations are summarized in Table~3. Initial masses of
primaries cover an interval from 1.3 $M_{\odot}$ where the subphotospheric
convection is supposed to cease, down to 0.9 $M_{\odot}$, for which the MS
life time becomes comparable to the Hubble time. Initial masses of the
secondaries were selected to obtain binaries with mass ratio close to 1 and
close to 0.5. Evolutionary tracks of the models from Table~3 are plotted in
period-AM diagram (Fig.~4). Dotted lines show AM evolution of binaries
in a detached phase and during fast exchange phase whereas solid lines give
tracks in a contact phase. 

%----------------------- Fig. 4 start --------------------------------
\begin{figure}
\begin{center}
\rotatebox{0}{\scalebox{0.45}{\includegraphics{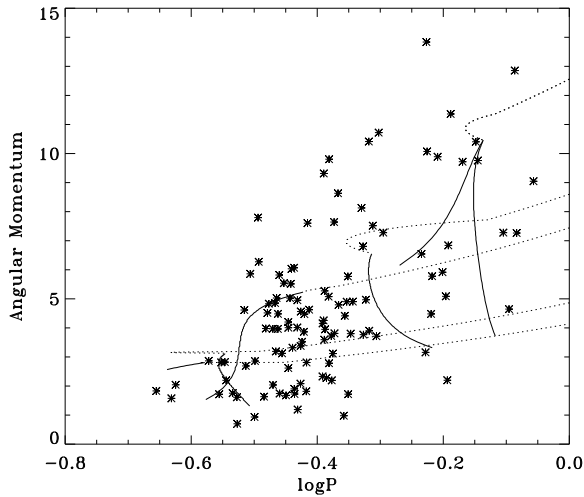}}}
\caption{\label{fig4}
The angular momentum distribution of 112 contact binaries. AM is in cgs
units $\times 10^{51}$. Evolutionary
tracks of model binaries listed in Table~3 are also shown. 
Parts of the tracks plotted with dotted lines correspond to
pre-contact phases and those plotted with solid lines describe binary 
evolution in contact.}
\end{center}
\end{figure} 
%----------------------- Fig. 4 end ---------------------------------

Two different models are listed for the binary 1.3+1.1 $M_{\odot}$. After the binary
was evolved through the detached phase and the fast exchange phase, two
different paths of evolution in contact were considered.  The first model
was evolved with mass transfer rate of $3.50\times 10^{-10} M_{\odot}$/year
and the second model with mass transfer rate of $3.75\times~10^{-10}~M_{\odot}$/year 
i. e. 7 \% higher. Such a small difference resulted in
distinctly different evolutionary tracks in the period-AM and period-mass
diagrams (see Figs.~4 and 5).  The lower mass transfer rate resulted in a
significant shrinking of the orbit followed by the overflow of the critical
Roche surface by the present primary when $q = 0.23$, i. e. still quite far
from the critical value of 0.1.  The slightly higher mass transfer rate
kept the orbit wide enough so that the contact configuration could exist
for a longer time and lose more AM. The overflow occurred when the mass
ratio approached the value of 0.1. This is an illustration of a great sensitivity of
contact binary evolution to the mass transfer rate, which points out to the
existence of a self-regulating mechanism of mass transfer with a negative
feed-back. There is no reason to assume that the mass transfer rate in real
binaries is constant. On the contrary, the rate is most likely adjusted
instantaneously to the evolutionary expansion rate of the secondary coupled
with the changing orbit. Fast expansion of the secondary results in an
increase of the mass transfer rate leading in turn to a widening of the
orbit and even faster increase of the Roche lobe (in spite of AML) which
cuts the rate down.  Slow expansion results in low mass transfer rate and
shrinking of the orbit because AML prevails. The shrinking secondary Roche
lobe enhances mass transfer which counteracts the decrease of the
orbit. The mass transfer rate adjusts itself to the evolutionary changes of
stellar radii and orbital parameters. Our constant mass transfer rates are
most likely equivalent to actual mass transfer rates averaged over the
whole contact phase.

%----------------------- Fig. 5 start --------------------------------
\begin{figure}
\begin{center}
\rotatebox{0}{\scalebox{0.45}{\includegraphics{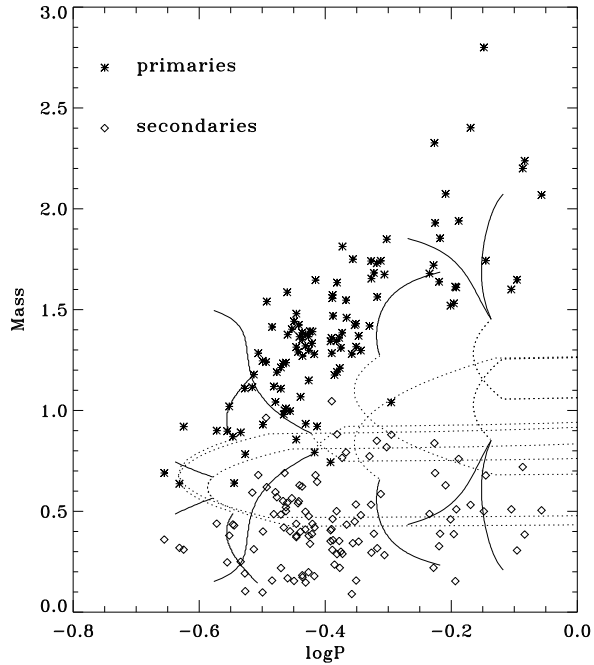}}}
\caption{\label{fig5}
Evolutionary
tracks of model binaries listed in Table~3. 
Their parts plotted with dotted lines correspond to
pre-contact phases and those plotted with solid lines describe binary 
evolution in contact. Over-plotted are observed masses of contact binaries
from Tables~1 and 2.}
\end{center}
\end{figure} 
%----------------------- Fig. 5 end ---------------------------------

The model calculations suggest that the newly formed contact binaries
appear near diagonal in the period-AM diagram (Fig.~4). It takes about 4-5
Gyr for progenitors of massive W~UMa stars to enter the contact phase. This
time rapidly increases with decreasing component masses up to about 12 Gyr
for progenitors of the least massive ones.  Subsequently, a contact binary
moves over next 1-2 Gyr downwards i. e. towards low AM. High- and
medium-mass binaries finish evolution in contact as extreme mass-ratio
binaries similar to the system AW~UMa in which the secondary has already
built a noticeable helium core \citep{Paczynski2007}. Such a binary is
soon expected to go coalescence into a rapidly rotating blue straggler
or a single and fast rotating star, possibly a giant of FK Com
type. Low-mass binaries have different evolutionary history as our models
show. After spending 12-13 Gyr in a detached configuration, they form a
tight, short period contact binary with very low AM. Evolutionary expansion
of the secondary in such a system is much slower than its counterparts in
more massive binaries, which results in a very low mass transfer rate. AML
caused by the winds probably dominates, shortening the period.  The binary
moves nearly horizontally in the period-AM diagram, as shown in Fig.~4
until both components overflow the critical Roche lobe and merge together.
Mass ratio hardly varies during the contact phase and does not reach any
extreme value, such as 0.1. 

Summarizing, we see that high mass systems may remain A-type
binaries over their whole evolution in contact. They evolve nearly
vertically in the period-AM diagram and finish evolution as extreme mass
ratio binaries similar to the system AW~UMa. Low mass systems do not evolve
towards such a configuration. Instead, their orbits shrink until both
components overflow the outer critical surface and merge together. If they
are born as W-type systems they may remain such during their whole contact
evolution. Medium mass binaries are probably formed as W-type with mass
ratios around 0.3-0.5 and they evolve as massive contact binaries
i.e. towards the extreme mass ratio A-type systems, although early
coalescence of some of them cannot be excluded. Better models are needed to
follow more accurately evolution of individual contact binaries.

Evolutionary tracks of the models listed in Table~3 are also shown in the
period-mass diagram (Fig.~5).  Similarly as in Fig.~4, dotted lines show
mass evolution of the components of a binary in a detached phase and during
mass transfer whereas solid lines describe mass evolution in a contact
phase. The lines run in pairs with symmetric shapes; upper branches
correspond to primaries and lower branches to secondaries. The models
reproduce correctly the observed properties of the binary components
described in \citet{GN2006}.

Evolutionary models presented in this paper do not include the problem of
energy transfer between the binary components. It is simply assumed that the
energy transfer does not influence stellar radii. \citet{Lucy1968} assumed that
energy is transferred by turbulent convection so that convective envelopes
of both components are on the same adiabat (i.e. the adiabatic constants of
both envelopes are equal). Recently, however, new models of energy transfer
have been developed in which the convective zone of the secondary remains
separated from a thin common envelope, extending above the inner critical
surface, by a radiative layer. Energy is transported by large scale
circulations in the common envelope \citep{MD1995, Kahler2002, Kahler2004}. 
The models do not violate the second law of thermodynamics (that was
a weakness of earlier models challenging the Lucy proposition) and both
components can be in thermal equilibrium \citep{Kahler2004}.  Neither model
of energy transfer, by turbulent convection or by large scale circulations,
can produce a secondary hotter than primary. Very efficient transport can
at most equalize both temperatures but lower efficiency results in hotter
primary.
In other words, A-type binaries are easily reproduced but not
W-type. Additional phenomena, like dark or hot spots, distributed on one or
both components, are invoked to explain W-type phenomenon.
Additional
arguments for the existence of such spots come from recurrent observations
of light curves of W~UMa stars, which show significant variability of
shape, minima depths and/or maxima heights. The most impressive example of
such variability is shown by the star OGLE BUL-SC27-506 in which light
curves taken over three consecutive seasons differ profoundly from one
another in shape and average brightness  (\citet{RP2002}).  The
season to season variations of the light curve have an amplitude comparable
to the depth of the minima.  Another example of such variations (although
on a much smaller scale) is presented by V839~Oph  (\citet{Gazeasetal2006}).  We
can only speculate, based on the observed data, that spot activity seems to
be weak in massive contact binaries, hence all they are of A-type, whereas 
low mass binaries are very active, hence they all are of
W-type. Binaries with intermediate masses can show either A-type or W-type
phenomenon or, sometimes, both in turn, as it happens on V839~Oph.

There is a group of high AM - short period stars lying above diagonal 
in Fig.~4 that are apparently not
covered by the computed models. They have AM of the order of $10^{52} gcm^{2}s^{-1})$
and periods around 0.4-0.5 d. Inspection of data from Tables~1 and 2 shows that these
are massive A-type contact binaries with tight orbits and with the
primaries significantly {\em undersized} compared to MS stars. Examples are
NN Vir, V351 Peg or AQ Tuc. Their present total masses are higher than the
total limiting mass for binaries with both components active. Assuming that
the derived parameters are correct, they must have lost AM via another
mechanism, not yet recognized, possibly similar to the one operating in hot
contact binaries. Our equations describing mass and AM evolution of
magnetically active stars are not applicable to such systems.

According to the data from Table~3, contact binaries with periods 0.5-0.7~d
have an age of about 5-6 Gyr but the age increases with decreasing period:
binaries with periods 0.3-0.4~d are about 9-10 Gyr old and binaries with
periods around 0.25 d are 12-13 Gyr old. Recently \citet{Bilir2005}
determined kinematic age of a large sample of field contact
binaries. Their results indicate that stars with periods 0.5-0.9~d are 3.2
Gyr old, stars with periods 0.4-0.5~d are 3.5 Gyr old, stars with periods
0.3-0.4~d are 7.1 Gyr old and those with periods 0.2-0.3~d are 8.9 Gyr old.  In
spite of the low number of our models, ages found from them are in a fair
agreement with these results. Ages of model binaries are about 20 \% higher
than kinematic ages but the steep trend of age with decreasing period is
well reproduced. In particular, the models suggest that not only extreme
mass ratio binaries but also the lowest mass binaries with short periods
and moderate mass ratios are approaching coalescence.  It is unfortunate
that because of apparent faintness the latter ones are observationally
neglected. Their accurate observations can shed light on the advanced
stages of evolution of old and/or evolved low mass binaries and
possibly on the formation of blue stragglers.

\section{Summary and Conclusions}

Analysis of the accurate observations of more than a hundred cool contact binaries reveals
the existence of several correlations among their physical and geometric
parameters. In particular, it is demonstrated that the knowledge of the
orbital period alone suffices to determine the absolute magnitude of the
system and masses and radii of the components with accuracy of about 15
\%. The primary components follow closely the mass-radius relation for main
sequence stars. The orbital AM increases on average with increasing period
but the correlation is rather poor because also the range of observed
values of AM increases rapidly with increasing period. 

Model calculations are presented according to scenario  suggested by
\citep{Stepien2004, Stepien2006a, Stepien2006b}. 
It is assumed that cool contact binaries are formed
from detached close binaries with initial (ZAMS) orbital periods of a
couple of days and total masses between about 1.4 and 2.6
$M_{\odot}$. Components of the binary lose mass and AM via magnetized
stellar wind which results in tightening of the orbit. The time scale of
orbital AML is of the order of several Gyr i. e. the same as the
evolutionary time scale of a more massive (primary) component. Both time
scales grow with decreasing stellar mass in a similar way, hence the
primary is at, or near TAMS when the shrinking Roche lobe reaches its
surface. RLOF results, followed by mass exchange between the components
through the common envelope phase. The model assumes that mass transfer
continues until mass ratio reversal and it stops only when the Roche lobe
of the hydrogen depleted, mass losing component becomes larger than the
stellar size. Depending on the detailed values of the involved parameters,
the other component (now more massive) may fill or under-fill its Roche
lobe. A contact binary is formed in the former case and a short period
Algol in the latter but, after an additional AML, it also converts into a
contact configuration. Both components are in thermal equilibrium. Details
of energy transfer between the components are not included into the
model. It is assumed that energy exchange takes place via large scale
circulations in the common envelope above the inner critical surface and
that it does not influence stellar radii.  As it was shown by
\citet{Kahler2004} both stars exchanging energy can, indeed, be in thermal
equilibrium. 
The evolution in contact is driven by a slow expansion of the presently
secondary component (which builds a helium core) followed by mass transfer
to the present primary component, accompanied by mass and AML due to
stellar winds. Depending on the relative importance of mass transfer and
AML an extreme mass ratio, or a very tight, medium
mass ratio binary will be formed. In either case both components merge
forming a single, rapidly rotating star. 

Precise duration of the contact phase depends on the adopted values of the
parameters influencing the evolution. Our results indicate that its typical
value is 1 - 1.5 Gyr (See Table~3). 
The average age of the contact binaries varies with
mass and orbital period, from about 5-6 Gyr for the most massive systems
with total mass of $\sim 2.5 M_{\odot}$ and periods of 0.5 - 0.7~d, up to
12-13 Gyr for the least massive systems with total mass of $\sim 1.2
M_{\odot}$ and periods around 0.25 d.

Detailed evolutionary paths of several binaries reproduce satisfactorily
the observed ranges of binary parameters except a few high mass A-type
systems with undersized primary components. They may belong to hot contact
binaries rather than cool, low mass contact binaries discussed in the
present paper.

\section{Acknowledgments}
\label{sect:Acknowledgments}
KS acknowledges the partial financial support of the Ministry of Science
and Higher Education through the grant 1 P03 016 28.

\end{document}